\PassOptionsToPackage{table}{xcolor}
\documentclass[conference,a4paper]{APSIPA2026}
\usepackage{amsmath}
\usepackage{graphicx}
\usepackage{multirow}
\usepackage{threeparttable}
\usepackage{booktabs}
\usepackage{xcolor}
\usepackage[backend=biber,style=ieee,maxnames=3,minnames=1]{biblatex}
\usepackage{flushend}
\usepackage{dblfloatfix}

\usepackage{geometry}
\usepackage{fancyhdr}

\fancypagestyle{firststyle}{\fancyhf{}
  \fancyhead[C]{2026 Asia Pacific Signal and Information Processing Association Annual Summit and Conference (APSIPA ASC)}}

\providecommand{\gc}{\cellcolor{gray!15}}   

\providecommand{\authorblockN}{\IEEEauthorblockN}
\providecommand{\authorblockA}{\IEEEauthorblockA}
\providecommand{\authorrefmark}{\IEEEauthorrefmark}

\begin{document}

\title{What Actually Makes Correlation-Based SSL Distillation Noise-Robust? A Mechanistic Correction}

\author{
\authorblockN{Fabian Ritter-Gutierrez\authorrefmark{1}\authorrefmark{2}, Nancy F. Chen\authorrefmark{2}, Eng Siong Chng\authorrefmark{1}}
\authorblockA{\authorrefmark{1}Nanyang Technological University, Singapore \qquad \authorrefmark{2}Institute for Infocomm Research (I2R), Singapore}
\authorblockA{E-mail: s220064@e.ntu.edu.sg, nancy\_chen@a-star.edu.sg, aseschng@ntu.edu.sg}
}

\maketitle
\thispagestyle{firststyle}
\pagestyle{empty}

\begin{abstract}
Self-supervised learning (SSL) speech models are accurate but large.
Knowledge distillation compresses them, but the student loses the teacher's
noise robustness. Correlation-based distillation addresses this with two
terms: cross-correlation aligning student and teacher's representations, and self-correlation
decorrelating the student's features. Both the original method and
De'HuBERT credited the self-correlation term without isolating it. We show
the opposite. On LibriSpeech-100
distillation with held-out CHiME-3 noise at 10\,dB, an unbiased
full-dimensional probe, a Pearson-variance decomposition, a same-noise
causal control, and a per-dimension analysis identify the
cross-correlation diagonal as the mechanism that encourages noise invariance, lowering
noise-classification accuracy from 76.98\% to 55.02\%, whereas adding the
self-correlation term back leaves it at 55.55\%. The self-correlation term
instead reorganises the feature space and improves accuracy on the clean
downstream tasks, but removes essentially no noise. Across nine speech and
music tasks this yields a concrete design rule for distillation: weight the
cross-correlation term for noise robustness, tune the self-correlation
term for better downstream performance, and sample teacher and student noise
independently.
\end{abstract}


\section{Introduction}

Self-supervised learning (SSL) speech models such as HuBERT~\cite{hubert}
and WavLM~\cite{wavlm} are accurate but large. Knowledge distillation
(KD) compresses them into deployable students such as
DistilHuBERT~\cite{distilhubert}. Compression, however, sacrifices the
teacher's robustness to noise~\cite{RobustDistilHuBERT}. Prior
noise-robust distillation methods augment the training data with noise but
remain sensitive to teacher robustness~\cite{RobustDistilHuBERT,
robustdistiller}. The prior method~\cite{ritter_noise_robust} replaced the
regression objective with a correlation objective adapted from Barlow
Twins~\cite{barlowtwins}, producing a noise-robust student regardless of
the teacher's noise robustness. De'HuBERT~\cite{ng2023hubert} applies a
similar correlation objective to make HuBERT itself noise-robust rather
than a distilled student.

Both correlation methods credit the same term for the robustness. The
prior method~\cite{ritter_noise_robust} states that robustness is
encouraged by minimising the student's self-correlation, and
De'HuBERT~\cite{ng2023hubert} attributes noise suppression primarily to
its self-correlation decorrelation term. Neither isolates the
self-correlation from the cross-correlation term. \cite{ritter_noise_robust}
asserts the attribution as a design rationale, and De'HuBERT supports it
only with a qualitative t-SNE and reports no term-isolating ablation. The
released noise-classification probe of the prior method is also
low-dimensional. It meanpools the representation, projects it onto 32
principal components, and classifies the noise type with a random forest.
We show that this PCA reduction distorts term-isolation comparisons by
exaggerating the apparent effect of the self-correlation term on noise.

A full-dimensional probe, a Pearson-variance decomposition, a same-noise
causal control, and a per-dimension analysis all point to the
cross-correlation term under independently sampled noise as the source of
the robustness. The self-correlation term reorganises the feature space and
improves clean downstream accuracy without removing noise. Without isolating each term's contribution, 
the published attribution would lead practitioners to emphasise the wrong loss component. 
The decomposition presented here makes that design choice interpretable.

This paper proposes no new loss; it reuses the distillation protocol of~\cite{ritter_noise_robust} and 
contributes a mechanistic correction of the causal attribution asserted in~\cite{ritter_noise_robust, ng2023hubert}, 
supported by four complementary diagnostics. Specifically:

\begin{enumerate}
\item A Pearson-variance decomposition predicting the cross-correlation
diagonal as the noise suppressor, active only under independent noise
(76.98\% to 55.02\% under the unbiased probe).
\item A probe-bias diagnosis showing the released PCA-reduced probe distorts
term-isolation comparisons, replaced here by a full-dimensional probe.
\item A same-noise causal control that disables the mechanism (55.55\% to
95.38\% under identical noise) and a per-dimension analysis confirming the
role separation.
\end{enumerate}

Additionally, we extend the evaluation to music and environmental-sound tasks to assess cross-domain generalisability.

\section{Background and Recap of the Method}

The distillation objective of~\cite{ritter_noise_robust} adapts Barlow
Twins~\cite{barlowtwins} from joint embedding to teacher-student
distillation. The student processes an input into a representation tensor of
shape $B\times P\times T\times D$ (batch, prediction head, time, feature
dimension). The frozen teacher produces a tensor of the same shape.
Correlation matrices are formed by an outer product over the feature
dimension $D$ with the time axis preserved, so no projector or extra
trainable parameters are needed. $\mathbf{C}_{\text{cc}}^{(p,t)}$ is the
batch-averaged outer product of the batch-normalised student and teacher
feature vectors. This is the matrix of Pearson correlations between student
and teacher dimensions. $\mathbf{C}_{\text{sc}}^{(p,t)}$ is the same
quantity computed on the student alone.

The cross-correlation loss drives the student-teacher correlation matrix
towards the identity, rewarding a unit diagonal and penalising off-diagonal
coupling. It is

\begin{equation}\label{eq:cc}
\mathcal{L}_{\text{CC}} = \sum_{p,t}\!\left[\sum_{i=1}^{D}\!\big(1-(\mathbf{C}_{\text{cc}}^{(p,t)})_{ii}\big)^2 + \lambda_{\text{cc}}\!\sum_{i=1}^{D}\sum_{\substack{j=1\\ j\neq i}}^{D}\!(\mathbf{C}_{\text{cc}}^{(p,t)})_{ij}^{2}\right]\!,
\end{equation}

where $\lambda_{\text{cc}}\!\ge\!0$ weights the off-diagonal penalty. The
self-correlation loss decorrelates the student's own feature dimensions by
penalising the off-diagonal entries of its self-correlation matrix. It is

\begin{equation}\label{eq:sc}
\mathcal{L}_{\text{SC}} = \lambda_{\text{sc}} \sum_{p,t} \sum_{i=1}^{D} \sum_{\substack{j=1\\ j\neq i}}^{D} (\mathbf{C}_{\text{sc}}^{(p,t)})_{ij}^{2},
\end{equation}

where $\lambda_{\text{sc}}\!\ge\!0$ controls self-decorrelation. DistilHuBERT~\cite{distilhubert} distils the teacher 
through prediction heads at layers 4, 8, and 12 by minimising an L1 and a frame-level cosine similarity loss $\mathcal{L}_{\cos}$. This original objective is
$\mathcal{L}{\text{KD}}$. The correlation method~\cite{ritter_noise_robust} keeps the same architecture but replaces L1 
with the two correlation terms, retaining the cosine term $\mathcal{L}_{\cos}$. $\mathcal{L}_{\text{KD}}$ serves as the no-correlation baseline. 
The full objective is

\begin{equation}\label{eq:cl}
\mathcal{L}_{\text{CL}} = \mathcal{L}_{\text{CC}} + \mathcal{L}_{\text{SC}} - \gamma\,\mathcal{L}_{\cos},
\end{equation}

where $\gamma$ weights the cosine term. The sections that follow test
whether the noise robustness originates in~\eqref{eq:sc} as previously attributed.

The training protocol follows~\cite{ritter_noise_robust}. Distillation runs
on clean LibriSpeech-100~\cite{librispeech} with additive noise from
MUSAN~\cite{musan} and WHAM!~\cite{wham} at signal-to-noise ratios (SNRs) in $[10,20)$\,dB. Noisy
evaluation uses held-out CHiME-3~\cite{chime3} noise at 10\,dB SNR. Two
augmentation regimes are studied. Setup~1 distorts the student input only,
and Setup~2 distorts both teacher and student with independently sampled
noise. The mechanistic analysis uses Setup~2 unless stated otherwise.
Downstream evaluation follows the SUPERB~\cite{superb} speech protocol as
in~\cite{ritter_noise_robust} and
the MARBLE~\cite{marble} music protocol as in~\cite{ritter_distill_speech_music} .

\section{The Pearson-Variance Mechanism}\label{sec:mech}

A short variance argument predicts which term in~\eqref{eq:cl} suppresses
noise. Under Setup~2 the teacher and student see the same clean speech with
independent additive noise. Write the teacher feature on dimension $i$ as
$h_i \approx s_i + n_{T,i}$ and the student feature as
$\hat{h}_i \approx s_i + n_{S,i}$, where $s_i$ is the shared speech
component and $n_{T,i}$ and $n_{S,i}$ are independent noise components.
Substituting these into the definition of the batch Pearson correlation and
using independence to drop the cross terms gives the diagonal entry of
$\mathbf{C}_{\text{cc}}$,

\begin{equation}\label{eq:pearson}
\rho(\hat{h}_i, h_i) = \frac{\mathrm{Var}(s_i)}{\sqrt{\mathrm{Var}(s_i)+\mathrm{Var}(n_{S,i})}\;\sqrt{\mathrm{Var}(s_i)+\mathrm{Var}(n_{T,i})}}.
\end{equation}

The teacher is frozen, so
$\mathrm{Var}(n_{T,i})$ is fixed during distillation. Maximising the
cross-correlation diagonal therefore requires minimising
$\mathrm{Var}(n_{S,i})$, so the diagonal term in~\eqref{eq:cc} implicitly
suppresses student noise variance. The self-correlation term
in~\eqref{eq:sc} does not enter~\eqref{eq:pearson}, so the algebra assigns
it no direct role in noise suppression.

This decomposition yields three testable predictions. First, the
cross-correlation diagonal is the suppressor, which Section~\ref{sec:probe}
tests with a factor decomposition. Second, independence is structurally
necessary. Identical teacher and student noise covaries, so the cross terms
no longer vanish and the diagonal would reward shared noise.
Section~\ref{sec:causal} tests this with a same-noise control. Third, the
effect scales with teacher noise variance, since a larger
$\mathrm{Var}(n_{T,i})$ makes the denominator more sensitive to
$\mathrm{Var}(n_{S,i})$. Section~\ref{sec:scope} relates this to the
choice of teacher.

\section{Term Isolation and Probe-Bias Diagnosis}\label{sec:probe}

Prior work attributed noise robustness to the self-correlation term without
isolating it from the cross-correlation term. We perform this isolation and
compare two noise classifiers to check for measurement bias. The student is
a two-layer transformer that produces 768-dimensional frame-level
representations. Probing and downstream evaluation of the distilled student use its
last-layer output. The first probe replicates the
procedure in the prior method's released code: mean-pooling over time,
PCA reduction to 32 of the 768 dimensions, and a random forest (PCA-32 RF).
The second is a logistic regression on all 768 dimensions without
dimensionality reduction (LR-768). Both probes classify which additive
noise type was applied to the audio signal. Lower accuracy indicates a more noise-invariant
representation.

For classification tasks we assess significance with McNemar's test on
paired per-sample outcomes. We use the continuity-corrected chi-square form
when the number of discordant pairs exceeds 25, and the exact binomial test
otherwise. Effect size is reported as Cohen's $h$ on the marginal
accuracies, with a 95\% confidence interval. For ASR word error rate we use
a paired bootstrap over utterances with 10{,}000 resamples. Within each
comparison family we control the false discovery rate with the
Benjamini--Hochberg procedure at $q=0.05$.

Table~\ref{tab:probe_comparison} shows that the two probes disagree. The
PCA-32 RF probe ranks the cross-correlation-only model
($\lambda_{\text{sc}} = 0$) at 27.07\%, far below the full
$\mathcal{L}_{\text{CL}}$, as though removing the self-correlation term
dramatically improves noise invariance. The LR-768 probe instead places the
cross-correlation-only model at 55.02\% and the full loss at 55.55\%, which
are essentially the same. The disagreement reveals a bias in the PCA-32
probe. PCA ranks directions by variance, not by class discriminability, and the
probe keeps only the 32 highest-variance directions. Without the
self-correlation term, the student's feature dimensions remain correlated,
and the noise-type information sits in low-variance combinations of them.
These combinations fall below the top-32 cut, so the projection reports
artificially low noise-classification accuracy even though the noise is
still linearly decodable by LR-768. The self-correlation term decorrelates
the feature space and aligns the feature axes with the principal axes. It
thereby moves the noise onto higher-variance individual dimensions that
survive the cut, and PCA retains more of it.
The full-dimensional probe discards no dimensions, removes this bias, and is
used for the remaining analysis.

\begin{table}[t]
\renewcommand{\arraystretch}{0.9}
\centering
\caption{Noise-classification accuracy (\%) under two probes. PCA-32~RF: PCA to 32 dimensions then Random Forest (100 bootstrap seeds, mean). LR-768: logistic regression on all 768 dimensions. Lower is more noise-invariant.}
\label{tab:probe_comparison}
\begin{tabular}{llcc}
\toprule
Model & Loss & PCA-32 RF & LR-768 \\
\midrule
DistilHuBERT & $\mathcal{L}_{\text{KD}}$ (clean) & 95.76 & 96.88 \\
Setup~1 & $\mathcal{L}_{\text{KD}}$ & 52.57 & 80.05 \\
Setup~1 & $\mathcal{L}_{\text{CL}}$ & 44.74 & 56.30 \\
Setup~2 & $\mathcal{L}_{\text{KD}}$ & 31.03 & 76.98 \\
Setup~2 & $\mathcal{L}_{\text{CL}}$ & 33.73 & 55.55 \\
Setup~2 & CC-only ($\lambda_{\text{sc}}\!=\!0$) & \textbf{27.07} & 55.02 \\
\bottomrule
\end{tabular}
\end{table}

Table~\ref{tab:mechanism_decomposition} decomposes the noise reduction one
factor at a time under the unbiased probe. Noise augmentation with the plain
distillation loss moves the LR-768 noise accuracy from the 96.88\% no-augmentation
baseline to 76.98\%, a drop of 19.9 points. This is a data-level
prerequisite, since a model never trained on noise cannot become invariant
to it. Switching to cross-correlation matching with the self-correlation
term disabled drops the accuracy further to 55.02\%, a reduction of 22.0
points, which is the dominant effect and matches the prediction
of~\eqref{eq:pearson}. Adding the self-correlation term back to obtain the 
full loss changes the accuracy by only $+0.53$ points to 55.55\%, which 
is negligible for noise. The decomposition identifies the 
cross-correlation diagonal as the source of noise robustness, 
while the self-correlation term plays a different role.

\begin{table}[t]
\renewcommand{\arraystretch}{0.9}
\centering
\caption{Decomposition of noise suppression (Setup~2, LR-768 probe). Each row changes exactly one factor while holding the others constant. Values trace to Table~\ref{tab:probe_comparison}.}
\label{tab:mechanism_decomposition}
\resizebox{\columnwidth}{!}{%
\begin{tabular}{lccl}
\toprule
Factor & LR-768 (\%) & $\Delta$ (pp) & Interpretation \\
\midrule
No noise aug.\ (baseline)            & 96.88 & ---      & noise fully visible \\
$+$ Noise aug.\ ($\mathcal{L}_{\text{KD}}$)        & 76.98 & $-$19.9  & prerequisite (data) \\
$+$ CC matching ($\lambda_{\text{sc}}\!=\!0$)      & 55.02 & $-$22.0  & \textbf{CC: dominant} \\
$+$ SC term (full $\mathcal{L}_{\text{CL}}$)       & 55.55 & $+$0.53  & SC: negligible \\
\bottomrule
\end{tabular}
}
\end{table}

The loss does not suppress every noise type equally.
Table~\ref{tab:per_class_noise} reports the per-class LR-768 accuracy, and we
read it through the reduction from the $\mathcal{L}_{\text{KD}}$ baseline to
$\mathcal{L}_{\text{CL}}$ rather than through the residual accuracies, which are
not comparable across classes.
The reduction follows a two-tier pattern. The loss strongly suppresses
independent additive noise, by 35.1 points for Gaussian and 31.2 points for
FSD50K. It barely moves reverberation and CHiME-3, by 8.4 and 9.1 points.
This pattern is consistent with~\eqref{eq:pearson}, whose suppression mechanism
acts on statistically independent additive noise and is therefore expected to
reach Gaussian and FSD50K most. Reverberation convolves the clean signal rather
than adding to it, and CHiME-3 is recorded noise with temporal structure and
partial correlation with the content; both depart from the additive model,
which may explain why the loss moves them least.

\begin{table}[t]
\renewcommand{\arraystretch}{0.9}
\centering
\caption{Per-class noise-classification accuracy (\%, per-class recall) under the LR-768 probe. Chance is 25\%.}
\label{tab:per_class_noise}
\resizebox{\columnwidth}{!}{%
\begin{tabular}{llcccc}
\toprule
Model & Loss & Reverb & CHiME-3 & FSD50K & Gaussian \\
\midrule
DistilHuBERT & $\mathcal{L}_{\text{KD}}$ (clean) & 100.0 & 97.2 & 90.4 & 100.0 \\
Setup~1 & $\mathcal{L}_{\text{KD}}$ & 92.9 & 78.7 & 66.1 & 83.0 \\
Setup~1 & $\mathcal{L}_{\text{CL}}$ & 80.4 & 54.6 & 41.3 & 48.7 \\
Setup~2 & $\mathcal{L}_{\text{KD}}$ & 89.3 & 75.0 & 65.5 & 78.2 \\
Setup~2 & $\mathcal{L}_{\text{CL}}$ & 80.9 & 65.9 & 34.3 & 43.1 \\
Setup~2 & CC-only ($\lambda_{\text{sc}}\!=\!0$) & 75.2 & 58.4 & 42.6 & 43.5 \\
\bottomrule
\end{tabular}
}
\end{table}

\begin{table*}[!t]
\setlength\tabcolsep{2pt}
\renewcommand{\arraystretch}{0.85}
\centering
\caption{Speech-task results from~\cite{ritter_noise_robust} under clean and noisy (CHiME-3, 10\,dB SNR) conditions. HuBERT/HuBERT+ are 95M parameters; distilled students are 23M. The lower block uses the HuBERT teacher under independent noise (Setup~2). Metrics: WER\%~$\downarrow$ for ASR, Acc\%~$\uparrow$ otherwise. $\Delta$ is clean-to-noisy degradation (smaller is more robust). Shaded row uses $\mathcal{L}_{\text{CL}}$.}
\label{tab:speech_background}
\resizebox{0.7\textwidth}{!}{%
\begin{tabular}{ll c | c c c | c c c | c c c | c c c}
\toprule
 &  &
 & \multicolumn{3}{c|}{\textbf{ASR} (WER$\downarrow$)}
 & \multicolumn{3}{c|}{\textbf{KS} (Acc$\uparrow$)}
 & \multicolumn{3}{c|}{\textbf{IC} (Acc$\uparrow$)}
 & \multicolumn{3}{c}{\textbf{ER} (Acc$\uparrow$)} \\
\cmidrule(lr){4-6} \cmidrule(lr){7-9} \cmidrule(lr){10-12} \cmidrule(lr){13-15}
Upstream & Dist. & Loss
 & cln & nsy & $\Delta$
 & cln & nsy & $\Delta$
 & cln & nsy & $\Delta$
 & cln & nsy & $\Delta$ \\
\midrule
\multicolumn{15}{l}{\textit{Baselines}} \\
HuBERT Base   & --   & --
 & 7.84  & 10.4  & 2.56
 & 96.30 & 93.74 & 2.56
 & 98.34 & 90.96 & 7.38
 & 64.92 & 57.69 & 7.23 \\
HuBERT+ Base  & --   & --
 & 8.08  & 9.41  & 1.33
 & 96.17 & 94.94 & 1.23
 & 98.37 & 97.02 & 1.35
 & 66.07 & 63.25 & 2.82 \\
DistilHuBERT  & --   & $\mathcal{L}_{\text{KD}}$
 & 15.83 & 29.65 & 13.82
 & 95.12 & 90.29 & 4.83
 & 91.20 & 66.71 & 24.49
 & 61.49 & 52.17 & 9.32 \\
\midrule
\multicolumn{15}{l}{\textit{Noise-robust distillation}} \\
DistilHuBERT  & Both & $\mathcal{L}_{\text{KD}}$
 & 16.98 & 20.47 & 3.49
 & 95.41 & 93.15 & 2.26
 & 93.03 & 89.29 & 3.74
 & 60.73 & 59.03 & 1.70 \\
\gc DistilHuBERT & \gc Both & \gc $\mathcal{L}_{\text{CL}}$
 & \gc 15.35 & \gc 18.18 & \gc 2.83
 & \gc 95.59 & \gc 94.38 & \gc 1.21
 & \gc 96.61 & \gc 94.06 & \gc 2.55
 & \gc 64.71 & \gc 63.09 & \gc 1.62 \\
\bottomrule
\end{tabular}
}
\end{table*}

\begin{table*}[!t]
\setlength\tabcolsep{2pt}
\renewcommand{\arraystretch}{0.85}
\centering
\caption{Cross-domain music and environmental-sound results under clean and noisy (CHiME-3, 10\,dB SNR) conditions. All metrics Acc\%~$\uparrow$; $\Delta=$ clean$-$noisy. The first row is vanilla DistilHuBERT without noise augmentation; the lower block is the HuBERT teacher under independent noise (Setup~2). Shaded row uses $\mathcal{L}_{\text{CL}}$. GenreID is omitted because the GTZAN benchmark is known to be unreliable.}
\label{tab:music_compressed}
\resizebox{0.7\textwidth}{!}{%
\begin{tabular}{ll c | c c c | c c c | c c c | c c c}
\toprule
 &  &
 & \multicolumn{3}{c|}{\textbf{SingerID}}
 & \multicolumn{3}{c|}{\textbf{VocID}}
 & \multicolumn{3}{c|}{\textbf{InstCls}}
 & \multicolumn{3}{c}{\textbf{ESC-50}} \\
\cmidrule(lr){4-6} \cmidrule(lr){7-9} \cmidrule(lr){10-12} \cmidrule(lr){13-15}
Upstream & Dist. & Loss
 & cln & nsy & $\Delta$
 & cln & nsy & $\Delta$
 & cln & nsy & $\Delta$
 & cln & nsy & $\Delta$ \\
\midrule
DistilHuBERT & -- & $\mathcal{L}_{\text{KD}}$
 & 68.04 & 36.41 & 31.63
 & 62.19 & 55.88 & 6.31
 & 61.42 & 22.75 & 38.67
 & 68.24 & 52.30 & 15.94 \\
\midrule
DistilHuBERT & Both & $\mathcal{L}_{\text{KD}}$
 & 57.84 & 42.21 & 15.63
 & 59.79 & 56.68 & 3.11
 & 60.64 & 33.45 & 27.19
 & 67.30 & 59.40 & 7.90 \\
\gc DistilHuBERT & \gc Both & \gc $\mathcal{L}_{\text{CL}}$
 & \gc 54.04 & \gc 41.46 & \gc 12.58
 & \gc 62.90 & \gc 58.43 & \gc 4.47
 & \gc 60.01 & \gc 37.85 & \gc 22.16
 & \gc 70.30 & \gc 64.69 & \gc 5.61 \\
\bottomrule
\end{tabular}
}
\end{table*}

\section{Same-Noise Control and Per-Dimension Structure Analysis}\label{sec:causal}

The analysis above decomposes each term's contribution but does not
establish causation. We now test the causal prediction that independent noise sampling is
necessary. Section~\ref{sec:mech} predicts that the mechanism requires
statistically independent teacher and student noise. We replace the
independent noise in Setup~2 with identical noise applied to both teacher
and student, and change nothing else.

\begin{table}[t]
\renewcommand{\arraystretch}{0.9}
\centering
\caption{Same-noise control. LR-768 noise-classification accuracy (\%). Chance is 25\%.}
\label{tab:same_noise_probe}
\begin{tabular}{llc}
\toprule
Model & Noise sampling & LR-768 (\%) \\
\midrule
DistilHuBERT (no noise aug.) & N/A & 96.88 \\
Setup~2 $\mathcal{L}_{\text{CL}}$ & Same & 95.38 \\
Setup~2 $\mathcal{L}_{\text{KD}}$ & Independent & 76.98 \\
Setup~2 $\mathcal{L}_{\text{CL}}$ & Independent & \textbf{55.55} \\
\bottomrule
\end{tabular}
\end{table}

Table~\ref{tab:same_noise_probe} confirms the prediction. With identical
noise, the full loss reaches 95.38\% LR-768 noise accuracy, close to the
96.88\% no-augmentation value. With independent noise the same loss reaches
55.55\%. Removing independence therefore disables the mechanism. A
behavioural check confirms this without any probe. Under identical noise,
intent classification (IC) accuracy degrades by 12.90 points from clean to
noisy (95.12 to 82.22), against only 2.55 points under independent noise
(96.61 to 94.06, Table~\ref{tab:sc_ablation}). Independent sampling
significantly outperforms identical sampling on IC by McNemar's test in
both conditions, with a much larger effect under noise. The downstream
collapse is consistent
with Equation~\eqref{eq:pearson} analysis.

\begin{table}[t]
\renewcommand{\arraystretch}{0.9}
\centering
\caption{Per-dimension noise analysis. Rows are Setup~2 models plus a no-augmentation reference (first row). LR-768 is total noise-classification accuracy. ``Dims $>$ 30\%'' counts the 768 feature dimensions individually above the 30\% threshold.}
\label{tab:per_dim_summary}
\resizebox{\columnwidth}{!}{%
\begin{tabular}{llcccc}
\toprule
Model & Noise & LR-768 (\%) & Max (\%) & Mean (\%) & Dims $>$ 30\% \\
\midrule
DistilHuBERT (no aug.) & N/A & 96.88 & -- & -- & 727 \\
$\mathcal{L}_{\text{CL}}$ & Same & 95.4 & 58.1 & 39.6 & 719 \\
$\mathcal{L}_{\text{KD}}$ & Indep.\ & 77.0 & 31.4 & 26.6 & 16 \\
$\mathcal{L}_{\text{CL}}$ & Indep.\ & 55.5 & 35.2 & 26.9 & 34 \\
CC-only ($\lambda_{\text{sc}}\!=\!0$) & Indep.\ & 55.0 & 30.5 & 26.3 & 5 \\
\bottomrule
\end{tabular}
}
\end{table}

\begin{figure}[t]
\centering
\includegraphics[width=0.92\columnwidth]{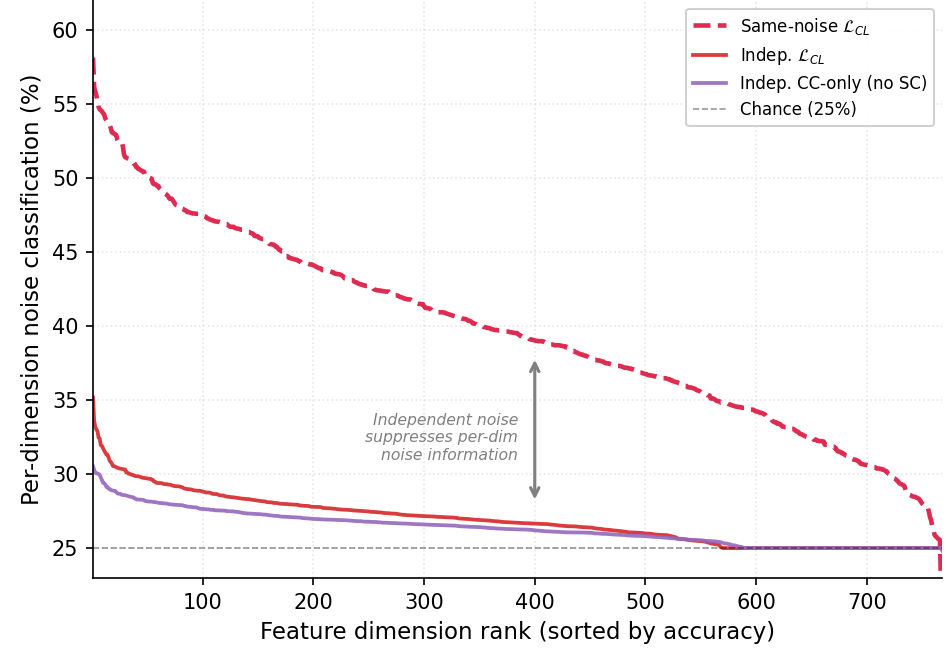}
\caption{Per-dimension noise-classification accuracy on the 768 feature
dimensions, ranked from most to least noise-informative (chance 25\%). At
matched total noise near 55\%, the full loss $\mathcal{L}_{\text{CL}}$
spreads noise over many more individually informative dimensions than the
cross-correlation-only model, while carrying the same total. Identical noise
saturates almost every dimension.}
\label{fig:perdim}
\end{figure}

The LR-768 probe used so far treats the 768-dimensional representation as a
single vector. A complementary per-dimension analysis trains 768 independent
classifiers, one per feature dimension, to reveal how noise is distributed
across individual dimensions (Table~\ref{tab:per_dim_summary}). Under
independent noise, the full loss and the cross-correlation-only model carry
nearly identical total noise (55.5\% and 55.0\%), yet differ in
per-dimension structure. The full loss spreads noise across 34 dimensions
that individually exceed the 30\% threshold (chance is 25\%), while the
cross-correlation-only model concentrates it in only 5. The
self-correlation term therefore does not remove noise but redistributes the
same total across more dimensions. Without the self-correlation term, the
same total noise resides in cross-dimensional correlations that no
single-dimension probe can detect but the joint LR-768 classifier can.
Under identical noise, the count rises
to 719 of 768, close to the 727 of the no-augmentation baseline,
confirming that the noise suppression mechanism is fully disabled.
Fig.~\ref{fig:perdim} shows the ranked per-dimension accuracy underlying
these counts.

\begin{table}[t]
\setlength\tabcolsep{2pt}
\renewcommand{\arraystretch}{0.9}
\centering
\caption{Self-correlation ablation (Setup~2, HuBERT teacher). CC-only sets $\lambda_{\text{sc}}\!=\!0$; full $\mathcal{L}_{\text{CL}}$ uses $\lambda_{\text{sc}}\!=\!5\times10^{-6}$. cln/nsy = clean/noisy.}
\label{tab:sc_ablation}
\resizebox{\columnwidth}{!}{%
\begin{tabular}{l|cc|cc|cc|cc}
\toprule
 & \multicolumn{2}{c|}{\textbf{IC} (Acc$\uparrow$)} & \multicolumn{2}{c|}{\textbf{KS} (Acc$\uparrow$)} & \multicolumn{2}{c|}{\textbf{ASR} (WER$\downarrow$)} & \multicolumn{2}{c}{\textbf{ER} (Acc$\uparrow$)} \\
Loss & cln & nsy & cln & nsy & cln & nsy & cln & nsy \\
\midrule
$\mathcal{L}_{\text{KD}}$ & 93.03 & 89.29 & 95.41 & 93.15 & 16.98 & 20.47 & 60.73 & 59.03 \\
CC-only ($\lambda_{\text{sc}}\!=\!0$) & 95.60 & 90.83 & 95.16 & 94.31 & 16.40 & 20.83 & 62.13 & 60.38 \\
Full $\mathcal{L}_{\text{CL}}$ & \textbf{96.61} & \textbf{94.06} & \textbf{95.59} & \textbf{94.38} & \textbf{15.35} & \textbf{18.18} & \textbf{64.71} & \textbf{63.09} \\
\bottomrule
\end{tabular}
}
\end{table}

The self-correlation term still matters, but for a different reason.
Table~\ref{tab:sc_ablation} ablates it on four speech tasks. Plain
distillation reaches 93.03\% clean IC accuracy, cross-correlation-only
reaches 95.60\%, and the full loss reaches 96.61\%. Emotion recognition
(ER) improves from 60.73 to 62.13 to 64.71 across the same three
settings. These gains appear even under clean conditions, where there is no
noise to remove, so the self-correlation term improves downstream accuracy
rather than suppressing noise. Downstream evaluation feeds the student's
frozen last-layer features to a lightweight linear head. Decorrelated
features provide more independent inputs to this head, a plausible
explanation for the observed accuracy gains. Relative to the
cross-correlation-only model, the self-correlation term yields significant
gains on IC and ER after Benjamini--Hochberg correction, while the
differences on KS and ASR are not significant.

Three independent lines of evidence converge on a consistent role
separation. The full-dimensional probe shows the self-correlation term adds
$+0.53$ noise-classification points. The per-dimension analysis shows
it redistributes noise across dimensions (34 vs 5) without changing the
total. The downstream ablation shows it instead improves clean task
accuracy. The noise-robustness mechanism is the cross-correlation diagonal
under independent noise. The self-correlation term organises the feature
space for downstream evaluation.

\section{Teacher Effect and Cross-Domain Generalisation}\label{sec:scope}

Equation~\eqref{eq:pearson} also explains a counter-intuitive
prior-work finding. A noise-sensitive teacher has a large
$\mathrm{Var}(n_{T,i})$, which makes the denominator
of~\eqref{eq:pearson} more sensitive to student noise and therefore
produces a stronger suppression gradient. The best student should then come
from the noise-sensitive teacher. The noise-robust comparison teacher,
HuBERT+, is HuBERT~\cite{hubert} after continual noise-augmented
pre-training~\cite{kuanpocontinualtraining}.
Table~\ref{tab:speech_background} is consistent with the prediction. With
the standard HuBERT teacher and the full loss, IC reaches 96.61\% clean and
94.06\% noisy, above the values reported with the HuBERT+ teacher in prior
work~\cite{ritter_noise_robust}. A noise-classification probe on the
teachers gives roughly 95\% for HuBERT against roughly 76\% for HuBERT+, so
HuBERT carries more noise variance. A layer-wise control on clean
LibriSpeech-100 finds the two teachers comparable in representational
quality, following the protocol
of~\cite{layerwiselivescu, livescu_all_models}. Canonical correlation analysis (CCA) similarity to
mel-spectrogram features is 0.630 against 0.640 at layer~4, with similar
mutual information with phone labels. The remaining difference is therefore noise variance.
Across the speech tasks in Table~\ref{tab:speech_background}, the full loss
improves over $\mathcal{L}_{\text{KD}}$ on IC, KS, and ER in both clean and
noisy conditions, all $p<0.02$ after Benjamini--Hochberg correction. The
effect sizes are small (Cohen's $h$ between 0.03 and 0.14), so the gains are
consistent rather than large. The ASR improvement is significant in both
conditions by the paired bootstrap.

To assess whether the noise-robustness mechanism transfers beyond speech,
we evaluate the HuBERT student under independent noise on music and
environmental-sound tasks (Table~\ref{tab:music_compressed}). On ESC-50
environmental sound~\cite{esc50}, the full loss degrades by 5.61 points
from clean to noisy (70.30 to 64.69) against 7.90 points for plain
distillation (67.30 to 59.40). Vocal identification improves similarly
(62.90/58.43 against 59.79/56.68). The ESC-50 improvement over
$\mathcal{L}_{\text{KD}}$ is significant ($p<0.01$ after correction); for
vocal identification the clean improvement is significant ($p=0.03$) while
the noisy improvement is not. The robustness mechanism therefore
carries over to non-speech additive noise. However, the augmentation
pipeline limits music tasks that depend on pitch and timbre cues. Singer
identification under the full loss reaches 54.04\% clean, well below the
68.04\% of vanilla DistilHuBERT, and instrument classification degrades by
22.16 points against 27.19 for plain distillation. The distillation pre-training pipeline
of~\cite{ritter_noise_robust} applies pitch-shift and band-reject
augmentations to encourage noise invariance for speech. These augmentations
remove pitch and timbre cues that music tasks depend on, so the same
pipeline that improves robustness on speech harms pitch-sensitive and
timbre-sensitive music tasks.

\section{Conclusion}

Correlation-based SSL distillation attributes its noise robustness to the
self-correlation term, yet neither the original method nor De'HuBERT
isolated this term from the cross-correlation term. We revisited this
attribution with four complementary diagnostics. A Pearson-variance
decomposition predicts the cross-correlation diagonal as the noise
suppressor under independently sampled noise, and a probe-bias diagnosis, a
same-noise causal control, and a per-dimension analysis all confirm it. The
self-correlation term does not suppress noise, it decorrelates the feature
space, producing more independent inputs for the linear downstream
evaluation, which explains the observed accuracy gains on clean tasks.

This yields a concrete design rule: weight the cross-correlation term for
noise robustness, tune the self-correlation weight for clean downstream
accuracy, and sample teacher and student noise independently. The mechanism
transfers to environmental-sound and vocal-identification tasks but not to
pitch-sensitive or timbre-sensitive music tasks, where the speech-domain
augmentation pipeline removes the cues those tasks need.
\printbibliography

\end{document}